# Highly tunable magnetic spirals and electric polarization in $Gd_{0.5}Dy_{0.5}MnO_3$

Chandan De[1], Rabindranath Bag[2], Surjeet Singh[2], Fabio Orlandi[3], P. Manuel[3], Sean Langridge[3], Milan K Sanyal[4], C. N. R. Rao[1], Maxim Mostovoy[5] and A. Sundaresan[1]

[1] *Chemistry and Physics of Materials Unit and International Center for Materials Science, Jawaharlal Nehru Centre for Advanced Scientific Research, Jakkur P.O., Bangalore 560064, India*

[2] *Indian Institute of Science Education and Research, Pashan, Pune 411008, India*

[3] *ISIS Facility, STFC Rutherford Appleton Laboratory, Chilton, Didcot, Oxfordshire, OX11 0QX, United Kingdom*

[4] *Saha Institute of Nuclear Physics, Bidhannagar, Kolkatta 700064 India*

[5] *Zernike Institute for Advanced Materials, University of Groningen, Nijenborgh 4, Groningen 9747 AG, The Netherlands*

Recent progress in the field of multiferroics led to the discovery of many new materials in which ferroelectricity is induced by cycloidal spiral orders. The direction of the electric polarization is typically constrained by spin anisotropies and magnetic field. Here, we report that the mixed rare-earth manganite, $Gd_{0.5}Dy_{0.5}MnO_3$, exhibits a spontaneous electric polarization along a general direction in the crystallographic *ac* plane, which is suppressed below 10 K but re-emerges in an applied magnetic field. Neutron diffraction measurements show that the polarization direction results from a large tilt of the spiral plane with respect to the crystallographic axes and that the suppression of ferroelectricity is caused by the transformation of a cycloidal spiral into a helical one – a unique property of this rare-earth manganite. The freedom in the orientation of the spiral plane allows for a fine magnetic control of ferroelectricity, i.e. a rotation as well as a strong enhancement of the polarization depending on the magnetic field direction. We show that this unusual behaviour originates from the coupling between the transition metal and rare-earth magnetic subsystems.

## Introduction

There has been a resurgence of interest in multiferroic materials showing the simultaneous presence of magnetic and ferroelectric orders, with a promising cross linking between them, which is an important parameter for applications such as magnetoelectric memory, spintronics, sensors and magnetoelectric microwave devices [1-10]. Coexisting magnetic and electric dipole orders are often found in frustrated magnets where the electric polarization is induced by unconventional spin orders breaking inversion symmetry of the crystal lattice [7-9]. Frustrated magnetism leads to many competing states and complex phase diagrams. In orthorhombic rare-earth ($R$) manganites $R$MnO$_3$ ($R$ = Dy, Tb and Gd), ferroelectricity is associated with the cycloidal spiral ordering of the Mn spins [1, 2, 11, 12]. The electric polarization vector lies in the spiral plane and is perpendicular to the spiral wave vector in accordance with the inverse Dzyaloshinskii-Moriya mechanism and the spin current model of the magnetically-induced ferroelectricity [13-15]. The orientation of the spiral plane depends on the ionic size of $R$-ion and applied magnetic fields [16]. For example, $R$MnO$_3$ with $R$ = Tb and Dy show the cycloidal spiral state with the wave vector along the $b$-axis and spins parallel to the $bc$ plane which induces a polarization, $P_c$, in the $c$-direction [17, 18]. In an applied magnetic field along the $b$-axis, the polarization vector re-orients from the $c$- to $a$-direction due to flop of spiral spins from the $bc$ to the $ab$ plane, in which the magnetic point groups change from *mm21'* to *2mm1'* [11]. GdMnO$_3$ is close to the borderline separating the nonpolar collinear A-type antiferromagnetic (AFM) state from the polar non-collinear spiral states [11]. Its precise location in the phase diagram remains a controversy and the nature of various competing phases is not well understood. Magnetization and single-crystal synchrotron x-ray data suggest that below 23 K the Mn spins order in the A-type AFM state which is weakly ferromagnetic and paraelectric [19, 20]. This is inconsistent with a small electric polarization observed in a narrow temperature interval around the Gd ordering temperature (~ 6.5 K) [11, 12]. Moreover, magnetic field applied along the *b*-axis induces a ferroelectric state below 23 K with the electric polarization, $P_a$, comparable to that in Tb and Dy compounds. In this state, both the Mn and Gd spins show a commensurate spin ordering along the *b*-axis [19, 20]. The field-induced ferroelectricity is believed to originate from a commensurate *ab* cycloidal spiral [20, 21]. Indeed, in a related compound, $Gd_{0.7}Tb_{0.3}MnO_3$, the *ab* spiral state with the polarization parallel to the *a* axis was observed in neutron diffraction studies [22].

In addition to the strong effect of the size of rare-earth ions on competing exchange interactions between $Mn^{3+}$ spins and thus the magnetic structure and dielectric properties of rare-earth manganites, the coupling between the *R*-ion and $Mn^{3+}$ magnetic subsystems also plays an important role. In the case of TbMnO₃, the electric polarization mainly comes from the Mn-magnetic subsystem. On the other hand, in DyMnO₃, an incommensurate magnetic ordering of Dy-ions, which is induced by the spiral ordering of Mn-spins and has the same wave vector, gives a significant contribution to the total electric polarization. This is evident from the large polarization drop below the commensurate magnetic ordering of Dy-spins ($T_N^{Dy} = 7$ K) when the rare-earth ions no longer contribute to the electric polarization and only the polarization induced by Mn-spins remains [23]. These facts prompted a number of systematic studies of orthorhombic manganites with mixed *R*-ions [24, 25], which revealed rich phase diagrams resulting from competition of collinear, non-collinear, commensurate and incommensurate magnetic phases that can coexist in spatially uniform and phase separated states. Nonetheless, multiferroic properties of manganites near the phase boundary between the *ab* and *bc* cycloidal phases as well as effects of the interplay between the Mn and $R$ magnetic orders are not well understood. An earlier study on polycrystalline $R_{0.5}Dy_{0.5}MnO_3$ ($R$ = Eu, Gd) showed an unusual behaviour of electric polarization and its enhancement under magnetic field [26].

In this work, we report magnetic and dielectric properties of $Gd_{0.5}Dy_{0.5}MnO_3$ (here after referred to as GDMO) single crystals along with a neutron diffraction study on polycrystalline $^{160}Gd$ enriched GDMO. The ferroelectric state of this material shows polarization both along the $a$- and $c$-axes, which is consistent with the fact that the cycloidal plane lays ~40$^o$ from the $ab$ plane, as revealed by neutron diffraction. Intriguingly, the appearance of short-range magnetic ordering of $R$-ions suppresses the ferroelectric cycloidal state of $Mn^{3+}$ spins and induces a helical spin structure with a small cycloidal component that results in a weak but finite electric polarization. The peculiar spins structure of GDMO allows for a fine control of the polarization direction in the ac plane that is not observed in any of the known $RMnO_3$ compounds. We will discuss this complex temperature and field behavior in terms of the magnetic symmetry and of the Mn and $R$-spins contributions.

## Results

**Magnetic phase transitions.** Figures 1(a) and (b) show the temperature dependence of heat capacity divided by temperature, $C/T$, and magnetization, $M(T)$, along the three crystallographic axes of a single crystal sample of GDMO. Heat capacity data show three anomalies that can be associated with the collinear incommensurate ordering ($T_{N1}^{Mn} = 39.5$ K), cycloidal ordering ($T_{N2}^{Mn} = 16.5$ K) of Mn spins and ordering of $R$-ions at $T_N^R = 5$ K, similar to the magnetic transitions observed in Tb and Dy manganites [1]. Magnetic susceptibility, dominated by rare earth spins, shows a cusp at $T_N^R$ (Fig. 1b) corresponding to a broad peak in $C/T$. The large suppression of the heat capacity peak at $T_N^R$ under a magnetic field $H_b = 8$ T and the nature of $M(H)$ data (inset of Fig. 1b) indicates the metamagnetic nature of $R$-ion ordering (see Supplementary Fig. 1-4 for M(H), dM/dH, $M_T$(H)) . On the other hand, the anomaly at $T_{N2}^{Mn}$ becomes more prominent increasing by 3 K at $H_b = 8$ T.

**Temperature-dependent spontaneous electric polarization**. First, we discuss the unusual behavior of spontaneous electric polarization in the temperature range $2 - 30$ K, measured after an electric poling (5.6 kV/cm) from 30 to 2 K. Figures 1(c) and (d) show the development of a spontaneous polarization both along the $a$ and $c$ directions ($P_a$ and $P_c$), respectively, below the cycloidal ordering temperature, which is consistent with the anomaly in the dielectric permittivities, $\varepsilon_a$ and $\varepsilon_c$, as shown in the respective figures. The observation of polarization along the $a$- and $c$- direction, in the single crystal specimens, indicates that the spontaneous polarization has a general direction in the $ac$ plane. Intriguingly, the polarization $P_a$ drops sharply at low temperatures from 1200 $\mu$C/m$^2$ at 7 K to a remnant polarization of ~ 100 $\mu$C/m$^2$ at 2 K, while $P_c$ drops from 200 $\mu$C/m$^2$ to 100 $\mu$C/m$^2$ in the same temperature interval. The decrease of electric polarization below 7 K indicates that the magnetic ordering of $R$-ions triggers a transition of the Mn magnetic subsystem into a non-ferroelectric (or a weakly ferroelectric) state. The presence of dielectric anomaly in both $\varepsilon_a$ and $\varepsilon_c$ at 5 K indicates that the low temperature state is a new polar phase (see Supplementary Fig. 5). The suppression of the ferroelectric spiral ordering of Mn spins by the $R$ magnetic ordering is unique to this compound. It may be mentioned here that the suppression of polarization reported in Gd$_{0.7}$Tb$_{0.3}$MnO$_3$ is due to transition from cycloidal to non-polar A-type antiferromagnetic phase [22].

**Effect of magnetic field on the polarization ($P_a$ and $P_c$)**. In the magnetic field applied along the $a$-direction, $P_a$ is suppressed while $P_c$ increases along with the recovery of the low temperature polarization (Fig. 2a,b). As in DyMnO$_3$, $\varepsilon_a$ and $\varepsilon_c$ show high sensitivity to the applied magnetic field near the flop transition (see insets of Fig. 2a,b) the so-called giant magneto-capacitance effect [2]. The ferroelectric Curie temperature increases by about 4 K under the magnetic field of 8 T. When the magnetic field is applied along the $b$ axis, both the Curie temperature and polarization ($P_a$ and $P_c$) increase monotonously with $H_b$ (Fig. 2c,d). At $H_b = 8$ T and 2 K, $P_a = 2500$ µC·m$^{-2}$, which is 25 times larger than the zero-field value and is comparable to the polarization in the $ab$ spiral state of DyMnO$_3$ [11]. While $H_b$ in TbMnO$_3$ and DyMnO$_3$ induces the flop transition from the $bc$-spiral to the $ab$-spiral state [2, 11], no flop transition is found in GDMO: remarkably, the polarization along the $c$-axis, $P_c$, also increases: at $H_b = 8$ T it is five times larger than in zero field (Fig. 2c). The behavior observed in $H_c$ is similar to that in $H_b$, except that the growth of $P_a$ is less dramatic (See Supplementary Fig. 6 for $P_a$, $P_c$ ($H//c$)). Furthermore, $P_a$ has a maximum around 3 T and decreases upon further increase of magnetic field. The polarization along $b$-direction is shown in Supplementary Fig. 7.

**Neutron diffraction results in polycrystalline $^{160}$Gd substituted GDMO.** In order to understand the unusual temperature and magnetic field dependence of spontaneous polarization, we have carried out neutron powder diffraction studies of $^{160}$Gd substituted GDMO. The refinement of the nuclear structure within the *Pbnm1'* space group returns good agreement factors between the observed and calculated data as shown by the Rietveld plots in Supplementary Fig. 8. The refinement of the average occupancy of the A site is in agreement with the nominal composition and details of the nuclear refinement are reported in the supporting information (Supplementary Table 1 and 2). Below $T_{N1}^{Mn} = 39.5$ K a set of new reflections, ascribable to magnetic ordering of the Mn sublattice, is observed and indexed with a propagation vector **k** = (0 0.308(4) 0). An attempt to refine the observed diffraction pattern with a spin density wave (SDW) aligned with one of the unit cell directions, gives poor results. For this reason the SDW axis was allowed to rotate in the *ab* plane and a good agreement with the data (see Supplementary Fig. 9) was obtained with the moment approximately aligned along the [120] direction with an amplitude of 2.83(9) $\mu_B$ at 25 K (see Supplementary Fig. 10). The magnetic ordering is described by a non-polar space group, point group *2/m1'* (a detailed description of the symmetry analysis is reported in Supplementary Table 3 and 4), in agreement with the polarization measurements. Upon crossing $T_{N2}^{Mn}$, a strong and abrupt increase of the magnetic intensity indicates a change in the spin configuration as also suggested by the pyrocurrent measurements and heat capacity data. It should be emphasized that we do not observe the appearance of new magnetic reflections and that all the measured satellites can be indexed with a single propagation vector k=(0 0.309(4) 0). This evidence combined with the observation of sharp magnetic reflections and a spontaneous electrical polarization in the *ac* plane in single crystal specimens suggest the presence of a single magnetic phase. As in the other *R*MnO$_3$ compounds [1], this transition is related to the appearance of a cycloidal spin structure. The refined magnetic structure is schematically shown in Fig. 3(a),(b),(c) and consists of a cycloidal spiral in the (102) plane with an ordered moment of 2.80(9) $\mu_B$ at 11 K, for a detailed symmetry description see Supplementary Table 5 and 6. The tilting of the manganese cycloidal plane by ~40 degrees from the *ab* plane is allowed by the magnetic point group *m1'* and it is in agreement with the observation of a spontaneous polarization in both the *a* and *c* directions. Moreover, assuming equal magnetoelectric coupling constants along the *a* and *c* directions, the induced polarization should be parallel to the [20-1] direction, in agreement with the hierarchy of the measured polarization along *a* and *c*, with $P_a > P_c$. The magnetic superspace group, *Pn1'(0β0)0s* (see Supplementary Table 5 and 6 for details), actually allows the cycloidal plane to freely rotate around the *b*-axis without symmetry breaking. This can explain the change of the polarization from $P_a$ to $P_c$, above $T_N^R$, when magnetic field is applied along the $a$ direction. Interestingly, the value of the propagation vector along the $b$ direction remains unchanged between the collinear and the cycloidal phases, in contrast with the other orthorhombic rare earth manganites.

Below 10 K a clear rise of diffuse scattering in the diffraction data is observed (see Supplementary Fig. 12) indicating the setting in of short range correlations. From the intensity of the diffuse scattering, heat capacity and magnetization measurements (see Fig. 1(a),(b)) it is possible to conclude that below $T_N^R$ a short range ordering of the rare earth moments develops. Moreover a new set of satellite reflections is observed, which can be indexed with the same propagation vector as the cycloidal structure. In particular, the observation of the 000+k and 110-k magnetic reflections, absent in the cycloidal phase, suggests the presence of a sizable magnetic component perpendicular to the propagation vector direction characteristic of a helical phase. The best refinement of the neutron data at base temperature is obtained within the same space group of the cycloidal phase (see Supplementary Fig. 13) and the resulting magnetic structure is drawn in Fig. 3(d), (e) and (f). The structure is characterized by a spiral spin arrangements with the spin rotating in the (041) and (04-1) planes alternating along the c-axis (see Fig. 3 (e) and (f)), with an ordered moment of 3.17(11) $\mu_B$. This spiral structure can be decomposed into a large non-polar helical component perpendicular to the propagation vector direction and a smaller polar cycloidal component. The magnetic structure and the point group $m1'$ are consistent with the pyrocurrent measurements indicating a strong decrease of the spontaneous polarization but still a finite value of $P_a$ and $P_c$. The finite value of the electrical polarization is due to the "cycloidal" part of the magnetic structure induced through the spin-current mechanism [13-15], whereas the large helical component perpendicular to the propagation vector is nonpolar since the parent space group (*Pbnm*) is not a ferro-axial group [27]. Owing to the larger helical component we will call this phase pseudo-helix (p-helix) below. It is worth stressing that the p-helix phase is still described in the same superspace group as the cycloidal one, since the magnetic symmetry do not constrain the phase relations between the different sinusoidal components, meaning that the phase transition between the two phases can occur with a continuous rotation or with a sudden jump of the cycloidal plane in an isostructural phase transition. Unfortunately, our diffraction data combined with the fact that the two phases have the same propagation vector, do not allow discriminating between the two cases. Figure 4 (left) shows the integrated intensity versus temperature of some magnetic reflections characteristic of each phase, highlighting the different transitions temperatures. It is interesting to note that the intensity of the 110-k and 000+k reflections, characteristic of the p-helix phase, increases below $T_{N2}^{Mn}$ following an almost linear dependence typical of a secondary (induced) order parameter. It is, therefore, likely that the rotation of the cycloidal plane from the (102) plane to the (041)/(04-1) is induced from the short range ordering of the $R$-ion moments on the A sites.

In order to study the role of the $R$-ions short range ordering of on the Mn spin and the strong increase in the electric polarization in the p-helix phase under external magnetic field, a series of in-field powder neutron diffraction experiments were performed. The increase of the applied magnetic field induces a clear decrease of the diffuse scattering with a concomitant increase of the diffracted intensity on some nuclear reflections as shown in Supplementary Fig. 14. These magnetic contributions, with the $k = 0$ propagation vector, are ascribable to a ferromagnetic polarization of the $R$-ion as also observed in the macroscopic magnetization measurements (Fig. 1(b)). As shown in Fig. 4, from the integrated intensity of the 110 reflection, the ferromagnetic polarization of the $R$-ions can be obtained both above and below $T_N^R$. More interesting is the effect of the applied field on the Mn sublattice ordering. Figure 4 (right) shows the integrated intensity of the 000+k and 001+k satellite reflections, characteristic of the p-helix and cycloidal phase respectively, versus the applied field above and below $T_N^R$. It clearly demonstrates that the application of an external magnetic field below $T_N^R$ induces the transition from the p-helix phase to the cycloidal one as indicated by vanishing 000+k satellite reflection, explaining also the strong increase of the polarization along the $a$ and $c$ axes. In contrast, the application of the magnetic field above the short range ordering of the $R$-ions does not change the cycloidal plane of the Mn spins, as shown by the integrated intensity of the magnetic satellites at 11 K (Fig. 4).

**Electric polarization in the p-helix phase.** The relatively small remnant polarization (~100 $\mu C/m^2$) is associated with the p-helix state. The presence of such a small polarization is demonstrated by our detailed pyroelectric current measurements as discussed below. The first evidence comes from pyrocurrent measurements with different poling intervals (Fig. 5(a)). For the poling intervals 8-2 K and 7-2 K, we see negative $(T_{N2}^{Mn})$ and positive $(T_N^R)$ pyrocurrent peaks corresponding to the appearance and disappearance of Mn polarization, respectively, similar to those observed for 30-2 K poling interval (the inset in Fig. 2c), except that the pyrocurrent values are smaller because the poling is done well below the Curie temperature. On the other hand, for the poling interval 6.5 - 2 K, we observe only a negative peak at $T_N^R$. This negative peak is not related to the disappearance of Mn polarization but indicates a polar state associated with the helical ordering. In order to further explore the polar nature of the low-temperature phase, we have measured pyrocurrent under various magnetic fields for the same poling interval. Figure 5(b) shows the suppression of the low-temperature polarization by magnetic field related to changes in the $R$-spin structure, which correlates with the meta-magnetic behavior shown in Fig. S1. The small negative pyrocurrent peaks at $T_{N2}^{Mn}$ in Fig. 5(b), corresponding to the Mn spiral ordering, are induced by the depolarization current at $T_N^R$. At $H_a = 8$ T, the low-temperature polar phase disappears and so does the peak at $T_{N2}^{Mn}$. It can be noted here that the polarization obtained by integrating the pyrocurrent in zero field (inset of Fig. 5(b)) is comparable to the remnant polarization at 2 K (Fig. 2). The ferroelectric nature of the low-temperature state ($T < T_N^R$) is further supported by the magnetic field dependence of the current measured at 2 K (Fig. 5(c)) showing a negative peak at ~ 1.1 T and a positive peak at ~1.5 T. These peaks correspond to disappearance of the low-temperature polarization and appearance of the spiral-induced polarization, respectively.

To further confirm the presence of small polarization in the low-temperature state, we have employed a different pyrocurrent (DC-bias) measurement [28]. In our measurement, the sample was cooled down to 2 K and the current $I_a$ was recorded in presence of an applied electric field (5.6 kV/cm) while warming in various magnetic fields ($H \| a$) [Fig. 5(d)]. In zero field, we observe a positive peak at 6 K followed by a small negative peak near 7 K, corresponding to the polarization and depolarization currents, respectively, which confirms the independent origin of ferroelectric order in the low-temperature state and is consistent with the dielectric anomaly at $T_N^R$ (Fig. 1(c), (d) and Supplementary Fig. 5). Similarly, we observe the polarization and depolarization currents due to the spiral ordering of Mn-spins at 10.5 K and 17 K, respectively. Under magnetic field, we see that the low-temperature polarization peaks are suppressed and only the peaks associated with the Mn spiral ordering remain.

**Discussion**

We summarize the temperature and magnetic field dependence of polarization in Fig. 6. For magnetic field applied along the $a$-direction, $P_a$ and $P_c$ at 10 K clearly demonstrate the switching of polarization direction from $P_a > P_c$ to $P_c > P_a$ related to rotation of cycloidal plane towards the $bc$ plane. At 2 K, $P_a$ is small because the cycloidal component in the p-helical ground state is small. The application of the field at 2 K induces the transition to the cycloidal phase, by destroying the short range AFM ordering of the R-ion, that will be in the same orientation at 10 K and hence with $P_c$ big and $P_a$ small. $P_c$ follows switching behavior at 10 K from $P_a \sim P_c$ to $P_c > P_a$. On the other hand, $P_a$ under $H//b$ shows a gigantic enhancement reaching a value (2500 $\mu C/m^2$) comparable to the maximum polarization reported for $DyMnO_3$. A schematic phase diagram is provided in Fig. 7 in order to show the various competitive magnetic phases with magnetic field and temperature.

**Theoretical model.** As the nature of the short-range rare earth magnetic order is unknown, we focus on Mn spin degrees of freedom, which we describe by the Landau free-energy expansion in powers of the A-type antiferromagnetic order parameter (ferromagnetic $ab$ layers coupled antiferromagnetically along the $c$ axis), $\boldsymbol{A} = (A_a, A_b, A_c)$. [29]:

$$f = \frac{c}{2} \boldsymbol{A} \left( \frac{d^2}{dy^2} + Q^2 \right)^2 \boldsymbol{A} + \frac{a}{2} A^2 + \frac{b}{4} A^4 + f_{ani}, \qquad (1)$$

where $a = \alpha(T - T_0)$ and $f_{ani} = \frac{1}{2} \delta a_a A_a^2 + \frac{1}{2} \delta a_c A_c^2 + \frac{1}{2} b_{ac} A_a^2 A_c^2 + \frac{1}{2} b'_{ac} \left( A_a \frac{dA_c}{dy} \right)^2$ is the anisotropy energy in which we keep only the most important terms. To describe effects of the magnetic rare-earth ions on the ordering of Mn spins we consider a strong temperature and field dependence of the anisotropy parameters, as discussed below. We then substitute $\boldsymbol{A} = (A_a, A_b, A_c)$ with $A_i = \Delta_i \cos(Qy + \varphi_i)$ ($i = a, b, c$) into equation (1) and minimize $f$ with respect to $\Delta_i$ and $\varphi_i$. The resulting phase diagrams for $H//a$ and $H//b$ together with the field-dependence of the electric polarization are shown in Fig. 8.

At high temperatures rare-earth spins are disordered and the easy axis anisotropy along the *b* direction present in manganites with both magnetic and non-magnetic *R*-ions favours the collinear incommensurate state, $A_b = \Delta_b \cos (Qy + \varphi_b)$ (brown color in Fig. 8 (a) and (d)). As temperature decreases and the modulation amplitude grows, the modulated $A_a$ and $A_c$ orders appear. For close values of $a_a$ and $a_c$, which is a special property of GDMO, we obtain the state with $\Delta_a, \Delta_c \neq 0$ and $\varphi_a = \varphi_c = \varphi_b \pm \frac{\pi}{2}$ (red color). This state, in which the *ab* and *bc* spirals coexist and have equal phases, is a tilted cycloidal spiral obtained e.g. by rotating spins in the *ab* spiral around the *b* axis. In our calculation $a_a < a_c$, which gives rise to a narrow strip of the *ab*-spiral phase between the collinear and tilted spiral phases.

As temperature decreases further, the coupling between the rare earth and Mn subsystems induces the *R*-order with the wave vector and symmetry of Mn spins, which in turn gives rise to an effective anisotropy of Mn spins with the Curie-like temperature dependence (see Supplement for details): $a_{a/c} = a_{a/c}^{(0)} - \frac{C_{a/c}}{T + \theta_{a/c}}$ . The decrease of $a_a$ and $a_c$ with decreasing temperature ultimately turns the *ac* plane into the magnetic easy plane, which triggers transition into the helical *ac* spiral state (violet color).

The polarization of the *R*-spins by an applied magnetic field, *H,* leads to their partial decoupling from the Mn spins. This effect is described by an increase of the 'Weiss constant', $\theta_{a/c}(H) = \theta_{a/c}(0) + \lambda C_{a/c} H^2$ $(\lambda, C_{a/c} > 0)$, in the susceptibility of the rare earth spins at the wave vector of the Mn spiral state, which affects the anisotropy parameters $\delta a_a$ and $\delta a_c$ and suppresses the *ac* spiral state (see Supplement for details). The difference in the effects of *H//a* (transition to the *bc* spiral state indicated by light green colour) and *H//b,c* (re-entrance into the tilted spiral state) results from different field-dependence of $a_a$ and $a_c$.

In conclusion, orthorhombic $Gd_{0.5}Dy_{0.5}MnO_3$ exhibits an electric polarization along a general direction in the *ac* plane which arises from the cycloidal spiral plane being rotated through ~40° from *ab* plane around the *b* axis. The cycloidal plane can be rotated by the application of an external magnetic field (*H//a*) inducing a rotation of the polarization direction in the *ac* plane. This process occurs without symmetry breaking, contrary to the spin flop transition reported in other known *R*MnO$_3$ compositions, and changes the $P_a/P_c$ ratio by more than an order of magnitude. Magnetic fields applied along the *b* axis result in a monotonous increase of both $P_a$ and $P_c$ with an unusually strong effect: $P_a(8\ \text{T})$ at 2 K is 25 times larger than $P_a(0\ \text{T})$. The short range rare-earth ordering suppresses the polar spiral ordering of Mn spins and gives rise to a largely helical state, not found in other rare-earth manganites, with a smaller electric polarization. Applied magnetic fields destroy the short-range ordering and lead to recovery of the cycloidal state.

**Methods**

**Experimental**. Single crystals were grown by the floating zone method and the phase purity of the grown crystal was confirmed by powder x-ray diffraction technique (Bruker D8 advance). For this purpose, small pieces were cut from the crystal boule and ground into fine powders. Crystal homogeneity was confirmed by optical microscopy under polarized light and scanning electron microscopy (Zeiss Ultra Plus, Germany). Three plate-like samples cut along the principle axes [100], [010] and [001] with a dimension of about 2 mm × 2 mm × 0.3 mm were used for measurements. Neutron diffraction measurements were made on polycrystalline sample containing [160]Gd isotope on the high-resolution powder diffractometer "WISH" at ISIS, UK [30]. The Jana2006 software were used to perform the Rietveld refinements of the neutron data whereas group theoretical calculation were carried out with the help of the ISODISTORT software [31, 32]. Magnetic measurements were performed in VSM SQUID. Heat capacity was measured in Physical Property Measurement System (PPMS). Dielectric constant measurement was performed using Agilent E4980A Precision LCR meter. Pyrocurrent was measured using Keithley Electrometer (6517A). Both the electrical measurements were performed in PPMS which provided the required temperature and magnetic field. For the pyroelectric current measurement, the sample was cooled to across the transition temperature with applied electric field. At the low temperature the electrodes were shorted. After waiting about 30 min, the electrometer was connected to the sample and the current was measured while heating the sample with 4 K/min rate without applying any electric field. For the DC-bias current measurement, the sample was cooled down to the lowest temperature without an electric poling and the current is recorded while warming the sample in presence of an electric field. This procedure enables to obtain both the polarization and depolarization currents in a single measurement at the transition temperature which is shown schematically in Supplementary Fig. 15 [33]. This technique has already been applied in several multiferroic studies [34-36].

**Acknowledgements** A. S. and C. De. acknowledge Sheikh Saqr Laboratory at Jawaharlal Nehru Centre for Advanced Scientific Research for providing experimental facilities. They thank the Nanomission council, Department of Science and Technology, India (SR/NM/Z-07/2015(G)) for the financial support and Jawaharlal Nehru Centre for Advanced Scientific Research (JNCASR) for managing the project. The authors acknowledge the Science and Technology Facilities Council for the provision of neutron beam time at the ISIS facility. MM acknowledges the NWO financial support.

**Figure captions:**

**Figure 1** Temperature dependence of heat capacity and magnetization. (a) and (b) show the heat capacity divided by temperature and magnetization as a function of temperature, respectively. Three anomalies corresponding to the Mn-spiral spin, cycloidal ordering and R-ion ordering at $T_{N1}^{Mn}$, $T_{N2}^{Mn}$, and $T_N^{RE}$ can be clearly seen. (c) and (d) show the electric polarization $P_a$ and $P_c$ and $\varepsilon_a$ and $\varepsilon_c$ versus temperature, respectively.

**Figure 2** Temperature dependence of electric polarization. (a) and (b) show the suppression and enhancement of the electric polarization of $P_a$ and $P_c$ when magnetic field is applied along $a$-direction. Temperature-dependence of $P_a$ and $P_c$ under applied field along $b$-direction (c) and (d). Insets show the corresponding dielectric data.

**Figure 3** Magnetic ordering of the different incommensurate phases of GDMO. Magnetic structures of the cycloidal phase (a), (b) and (c) and for the p-helix phase (d), (e) and (f) in different projections. (a) 3D view of the cycloidal phase: (b) projection of the cycloidal structure in the $ac$ plane showing the spins in the (102) plane; (c) projection along the [102] direction showing the rotation of the cycloidal structure in the (102) plane. (d) 3D view of the p-helix phase, (e) projection of the p-helix structure in the $ac$ plane showing the perpendicular component to k, (f) projection along the [041] direction showing the rotation of the Mn spins in the (041) plane. In all panels are reported the axes directions refer to the crystal structure (red $a$-axis, green $b$-axis and blue $c$-axis)

**Figure 4** Integrated intensity of magnetic reflections characteristic of each magnetic phase as function of temperature (a), (b), (c) and applied magnetic field (d), (e) and (f).

**Figure 5** Pyrocurrent, magnetoelectric current and DC-*bias* current. (a) Pyrocurrent data recorded along the *a*-axis for different poling intervals between 8 K to 2 K with 5.6 kV/cm. A negative peak appears at the rare-earth ordering when the poling field is applied from 6.5 K to 2 K. (b) Pyrocurrent measured along the *a*-axis under various $H_a$ after poling from 6.5 to 2 K. Inset shows polarization along the *a*-axis in zero magnetic field poled between 6.5 K to 2 K (c) magnetoelectric current recorded at 2 K while ramping the field from 0 to 3 T after poling the sample from 6.5 to 2 K with 5.6 kV/cm in zero magnetic field. (d) DC- bias current measured along the *a*-axis with 5.6 kV/cm at various magnetic fields along the *a*-axis.

**Figure 6** Magnetic field dependence of electric polarization. (a) Change of $P_a$ and $P_c$ under an applied field along *a*-direction and (b) Enhancement of polarization $P_a$ and $P_c$ under $H//b$.

**Figure 7** Experimental phase diagram of the GDMO system. Schematic phase diagram of various magnetic phases with temperature and magnetic field.

**Figure 8** Phase diagrams and magnetic field dependence of the electric polarization for the model (1). Phase diagrams for $H//a$ (a) and $H//b$ (d) showing the regions of the paramagnetic (white), incommensurate collinear with spins parallel to the *b* axis (brown), *ab* spiral (dark green), *bc* spiral (light green), the tilted *(a/c)b* spiral (red) and the *ac* spiral (violet) phases. The coefficient, *a*, of the Landau expansion of the free energy (1) plays the role of temperature. Also plotted is the corresponding temperature dependence of the electric polarization, $P_a$ (blue) and $P_c$ (red) in zero field (b), $H_a = 0.5$ (c), $H_b = 0.3$ (e) and $H_b = 0.8$ (f). See supplementary for details and the model parameters used in the calculation.

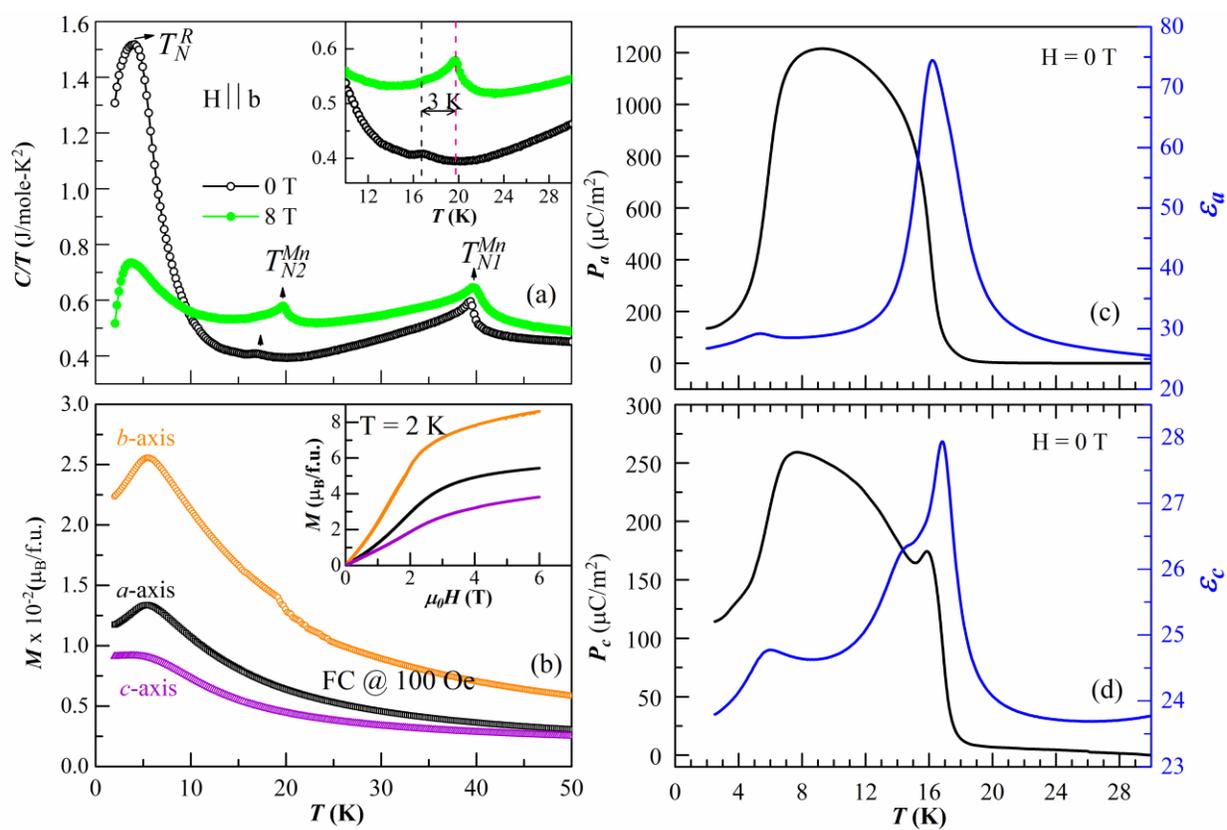

Figure 1

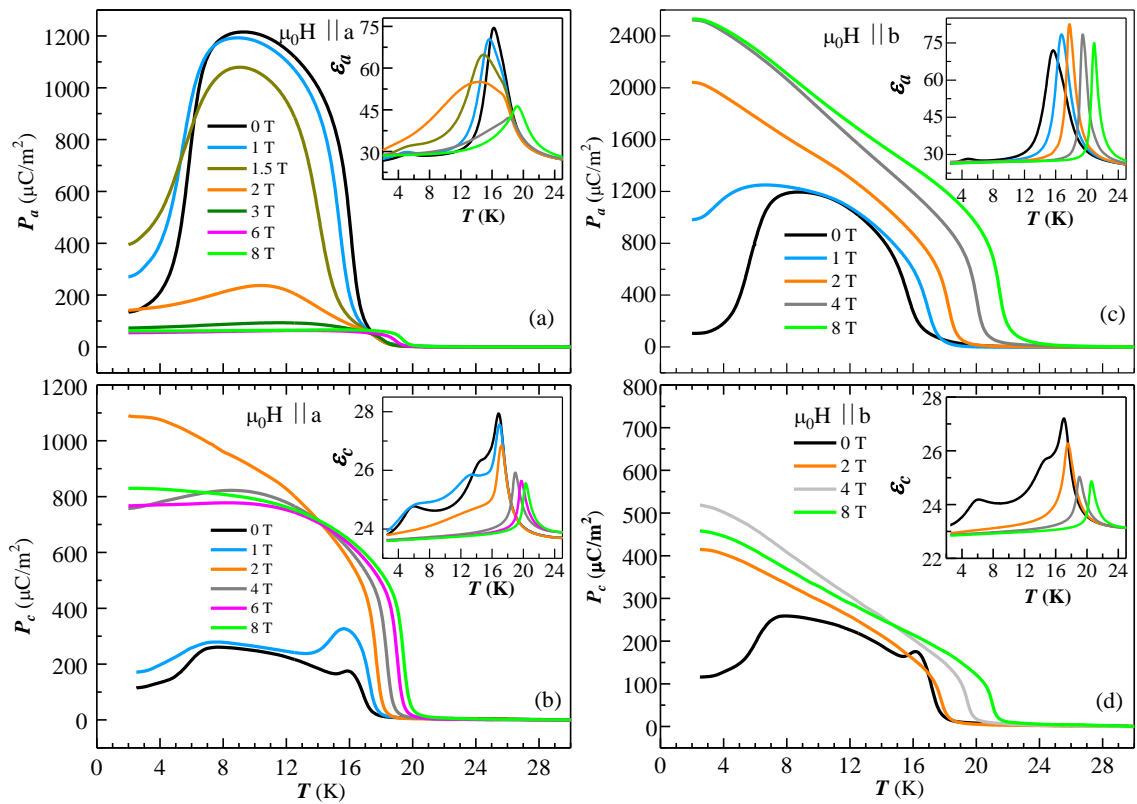

Figure 2

## Cycloidal phase

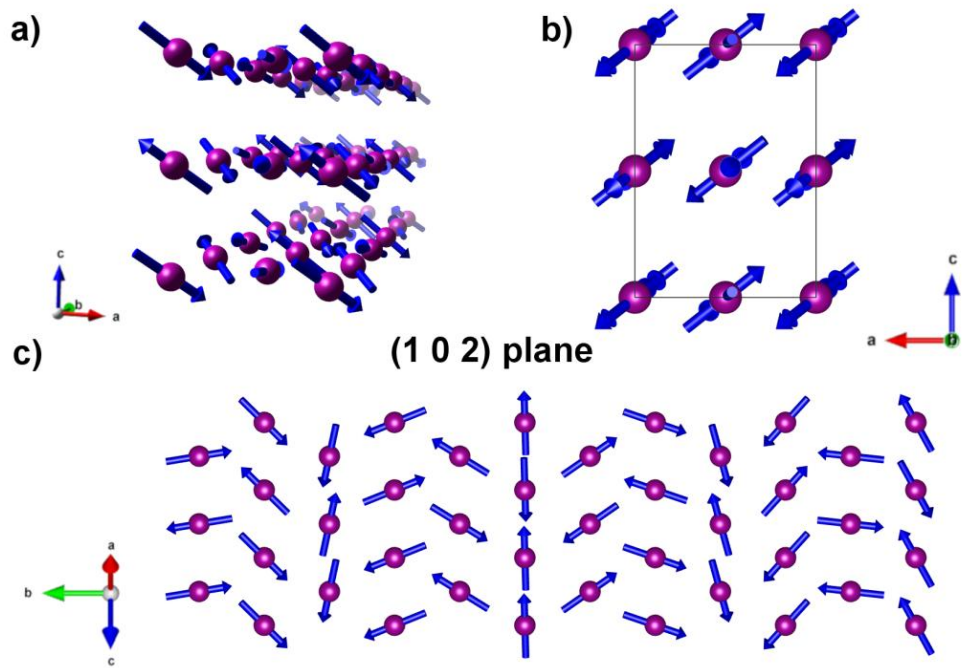

**(1 0 2) plane**

## Pseudo-helix  phase

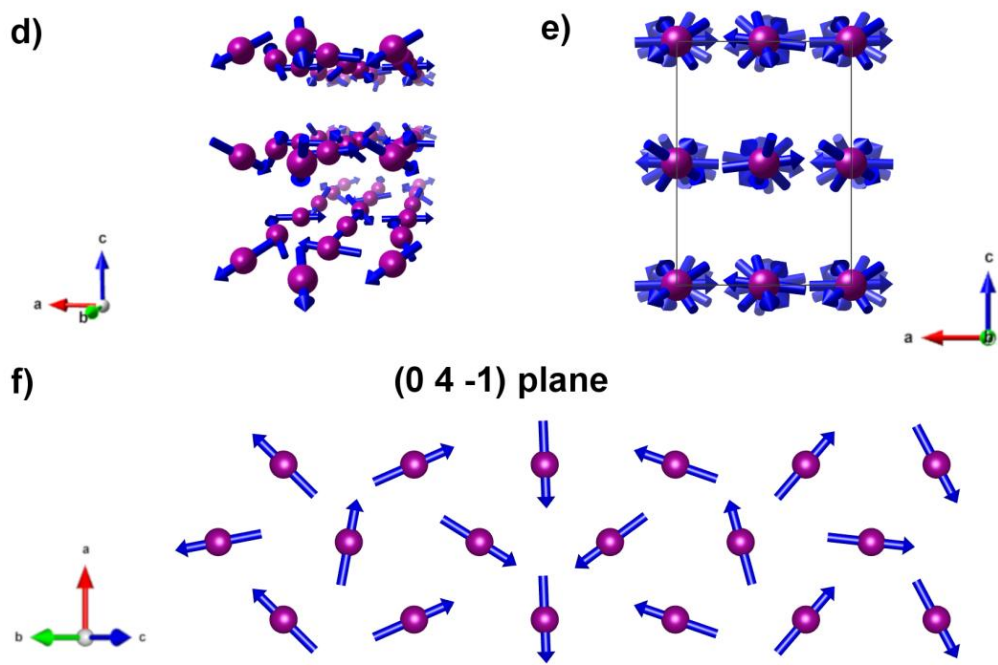

**(0 4 -1) plane**

Figure 3

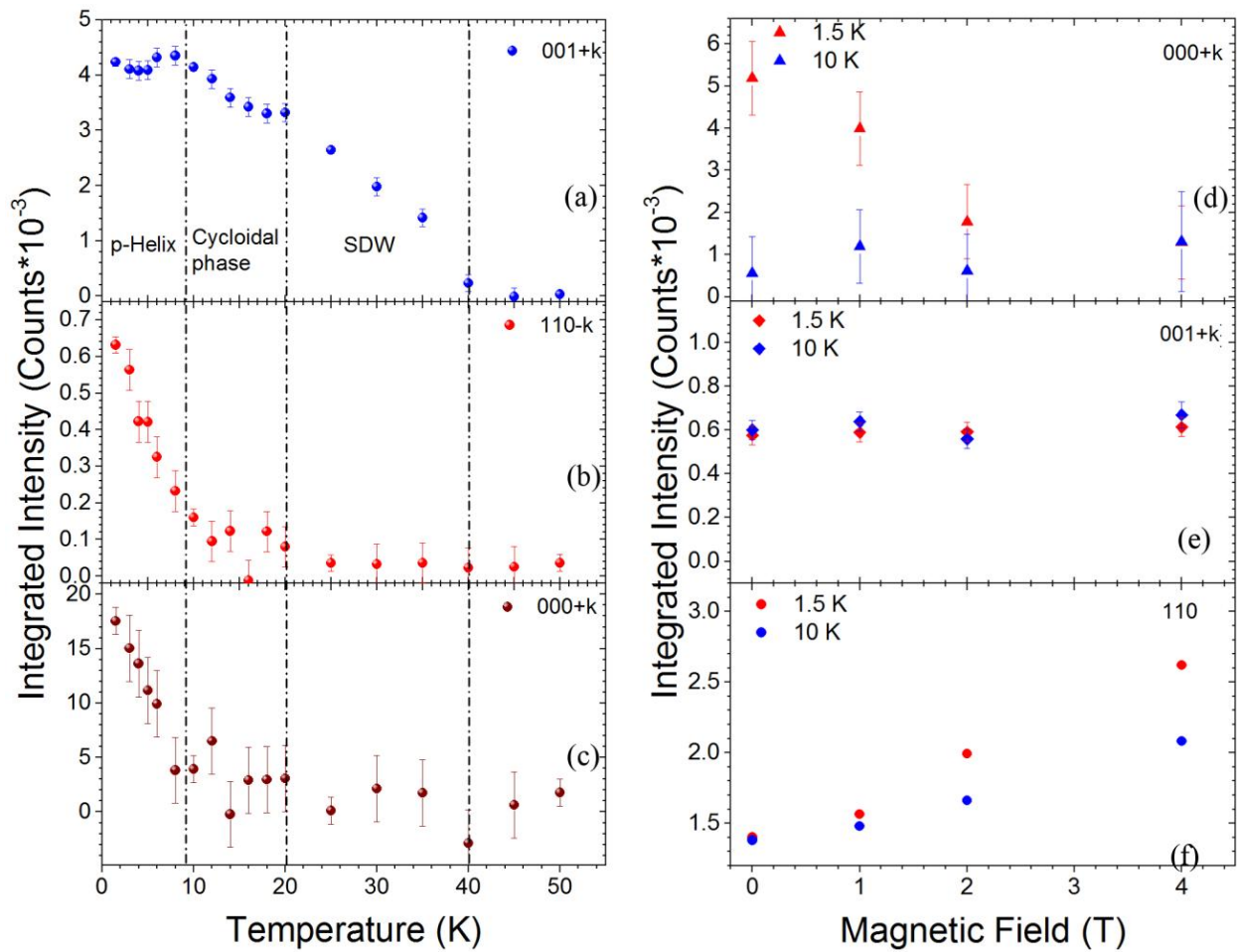

Figure 4

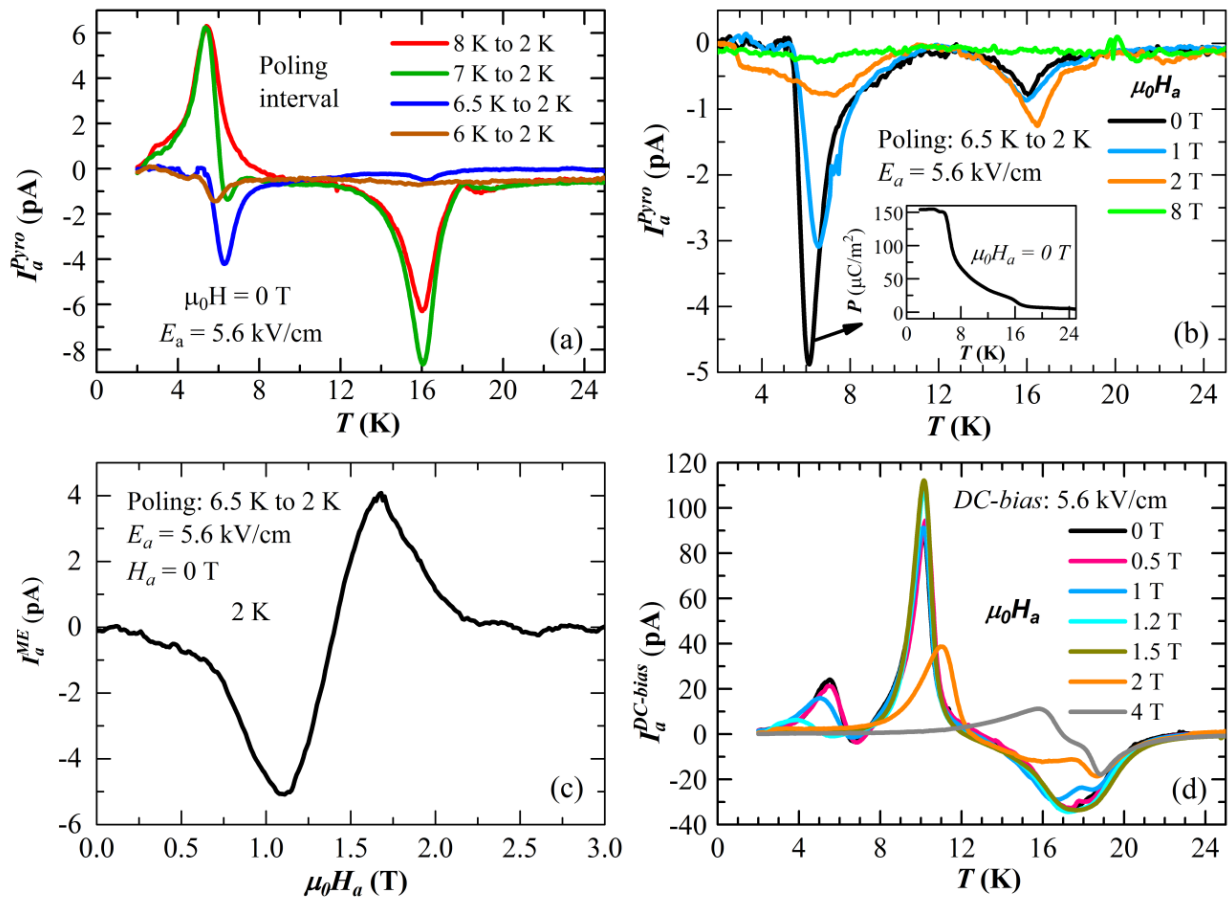

Figure 5

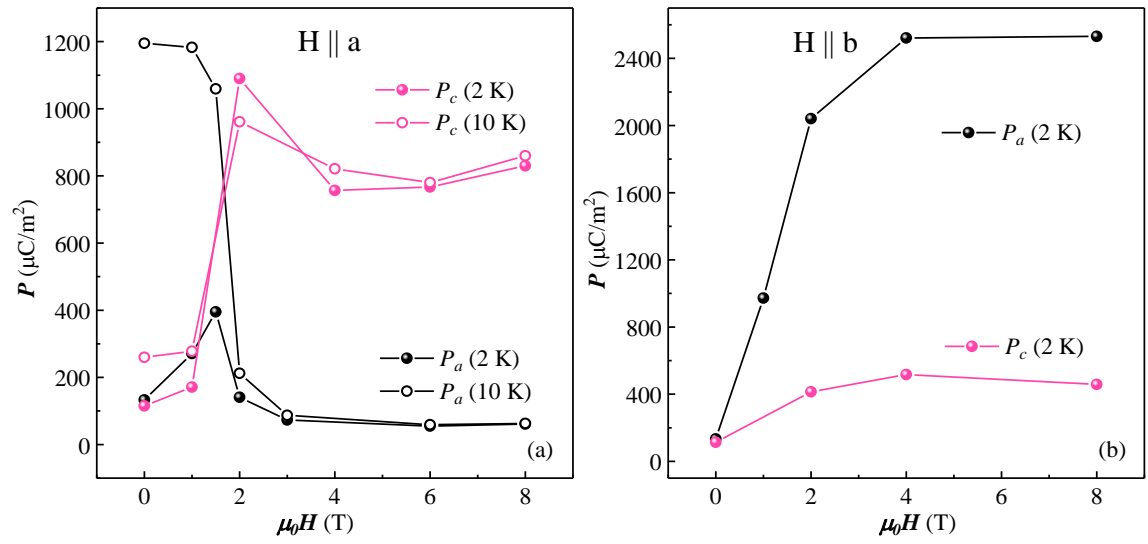

Figure 6

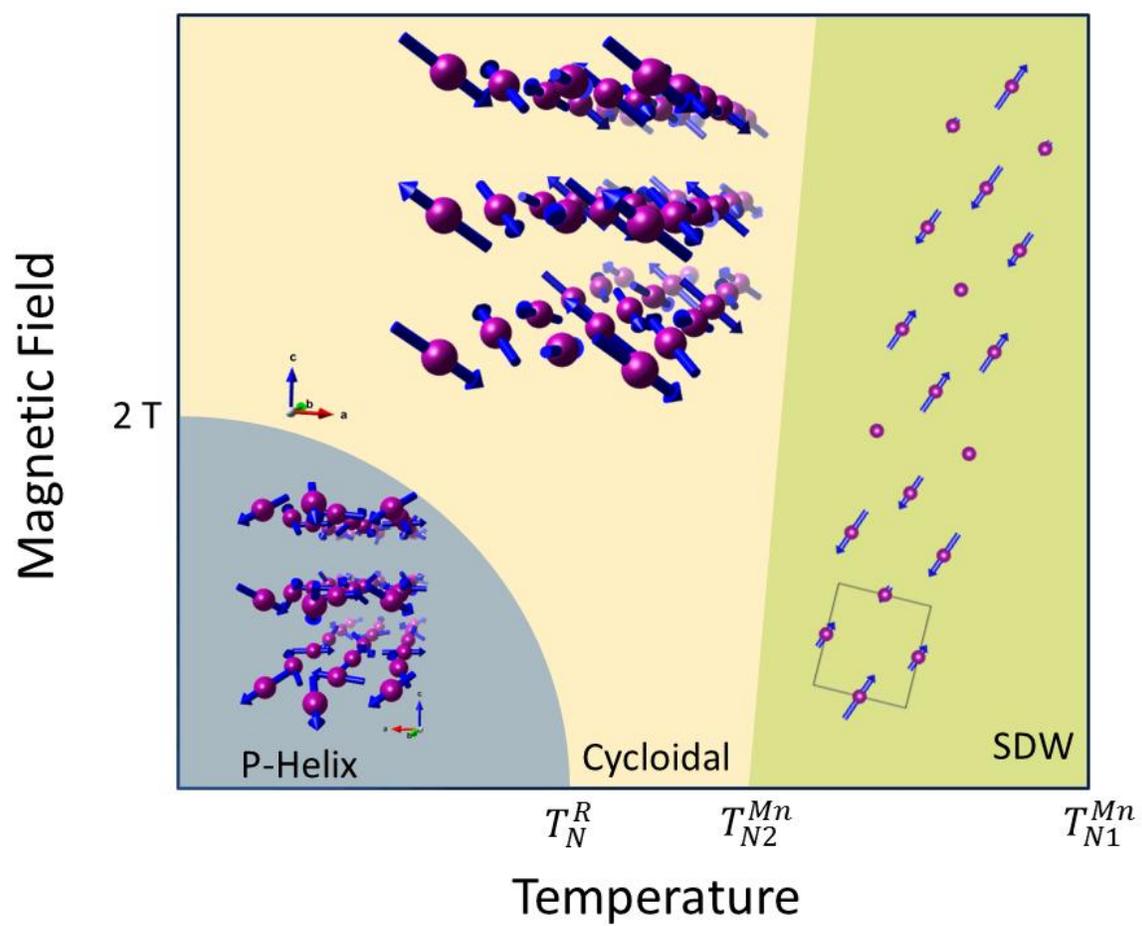

Figure 7

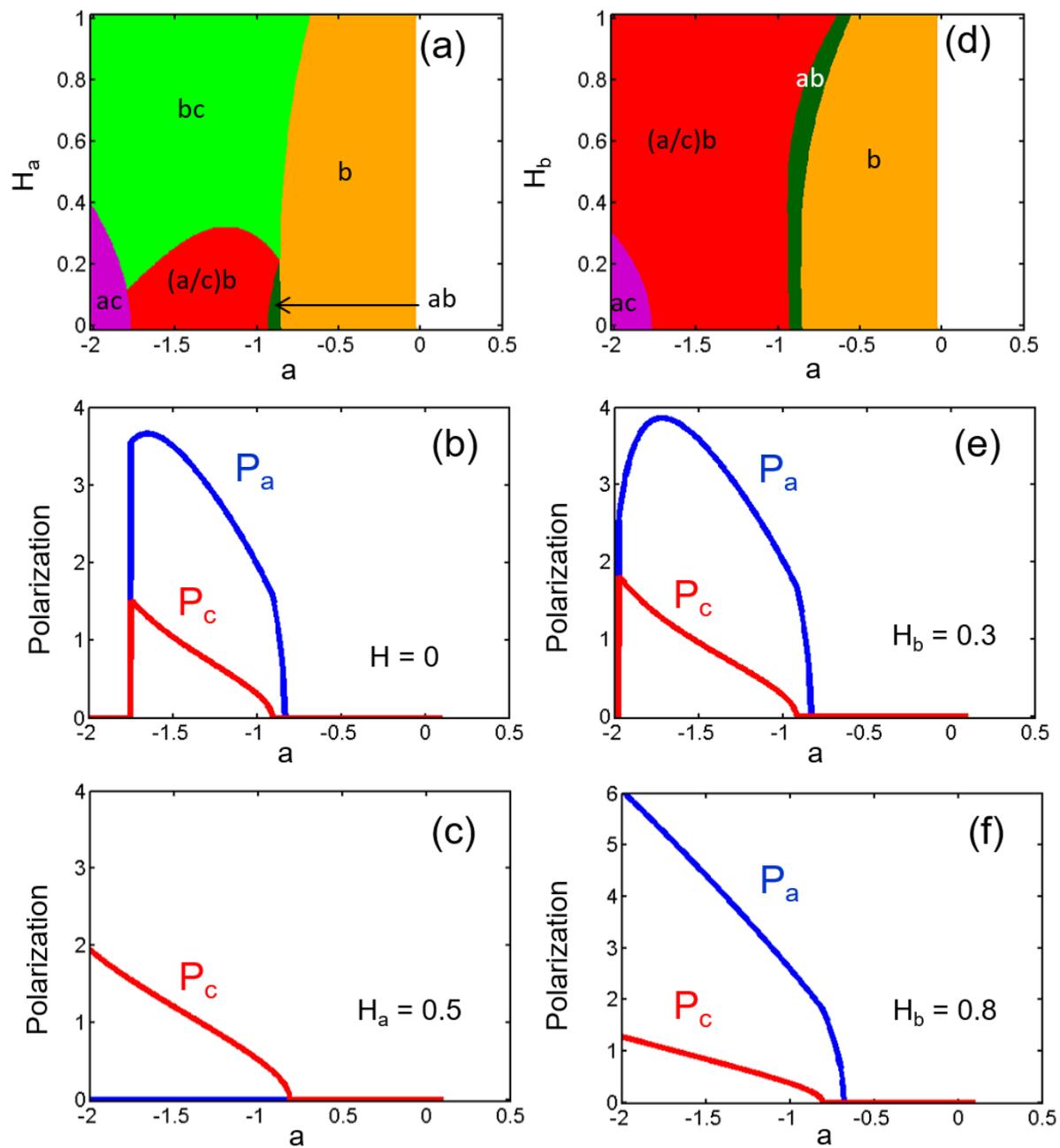

Figure 8